\documentclass[fleqn,usenatbib]{mnras}

\usepackage{newtxtext,newtxmath}
\usepackage[T1]{fontenc}

\DeclareRobustCommand{\VAN}[3]{#2}
\let\VANthebibliography\thebibliography
\def\thebibliography{\DeclareRobustCommand{\VAN}[3]{##3}\VANthebibliography}

\usepackage{graphicx}	% Including figure files
\usepackage{amsmath}	% Advanced maths commands
\title{Search for periodic radio emission of gamma pulsar J0357+3205}

\author[S. A. Tyul'bashev et al.]{
Sergei A. Tyul'bashev,$^{1}$\thanks{E-mail: serg@prao.ru (SAT)}
Marina A. Kitaeva,$^{1}$
\\
$^{1}$ P.N. Lebedev Physical Institute of the Russian Academy of Sciences, Astro Space Center, Pushchino Radio Astronomy Observatory,\\
Radiotelescopnaya 1a, Moscow reg., Pushchino, 142290, Russia \\
}

\date{04.07.2019}

\pubyear{04.07.2019}

\begin{document}
\label{firstpage}
\pagerange{\pageref{firstpage}--\pageref{lastpage}}
\maketitle

\begin{abstract}
The search for radio emission from the gamma pulsar J0357+3205 in the meter wavelength range was carried out. Periodic emission was found in one of the 1,700 observation sessions. The average pulsar profile is single-component with a half-width of 20-25 ms. Estimation of the pulsar flux density is 14 mJy.

\end{abstract}

\begin{keywords}
radio-quiet pulsars; observations at low frequencies; data analysis
\end{keywords}

%%%%%%%%%%%%%%%%%%%%%%%%%%%%%%%%%%%%%%%%%%%%%%%%%%

%%%%%%%%%%%%%%%%% BODY OF PAPER %%%%%%%%%%%%%%%%%%

\section{Introduction}
The launch of the Fermi telescope led to the discovery of a number of gamma pulsars (\citeauthor{Abdo2009}, \citeyear{Abdo2009}). The search for periodic radio emission from these pulsars has shown that in the radio range many of them have low luminosity (\citeauthor{Camilo2009}, \citeyear{Camilo2009}), that is, they are radio-quiet pulsars. In 2018, a conference was held in China, at which a report was presented with the presentation of the results of observations on the 500-meter FAST telescope of one of these pulsars (\citeauthor{PeiWang2018}, \citeyear{PeiWang2018}). According to the report, the periodic radio emission of gamma pulsar J0357+3205 was detected in FAST observations, and also confirmed by observations on the 300-meter Arecibo telescope.

Since 2013, the Large Phased Array (LPA) of Lebedev Physics Institute (LPI) radio telescope performed a daily survey of the sky at declinations $-9^o$ to $+55^o$. This survey is used, among other things, to search for pulsars in the summed power spectra (\citeauthor{Tyulbashev2017}, \citeyear{Tyulbashev2017}). Summation of the spectra makes it possible to increase sensitivity by increasing the total observation time for a given direction. Pulsar J0357+3205 falls into the study area, so it is possible to search for it in monitoring data for an interval of 5 years. Despite the high sensitivity in the search for summed spectra, there were no pronounced harmonics whose inverse frequency would correspond to the pulsar period (P=0.444104 c). On the other hand, it is known that pulsars have both internal (intrinsic) and external variability induced by the interstellar medium. Pulsar flare activity is also possible, or periods of long nullings. Therefore, it is likely to detect a pulsar in individual records, whereas it will not be visible in the total power spectra. At the moment, there are more than 1700 observation sessions in the direction of the pulsar J0357+3205. In this paper, we have attempted to search for its periodic radiation from all available data.

\begin{figure*}
\begin{center}
	\includegraphics[width=1.0\textwidth]{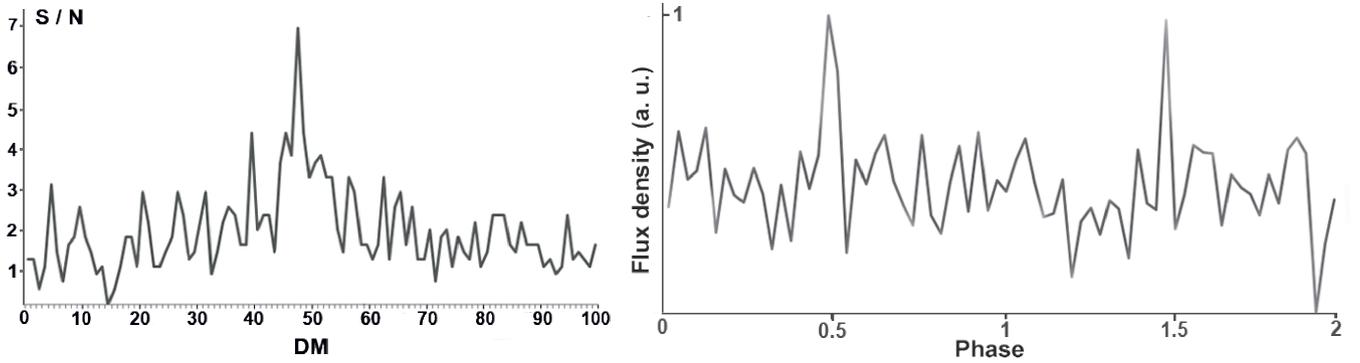}
    \caption{The left part of the figure shows the dependence of S/N on DM. The horizontal axis shows the dispersion measure being tested. The vertical axis shows the signal-to-noise ratio on the dispersion measure under test. It can be seen that the maximum dependence falls on the dispersion measures of 46-48 $pc/cm^3$. The right part of the figure shows the average profile of the pulsar J0357+3205 for December 23, 2016, accumulated with a double period. The pulse phase in the double profile is shown along the horizontal axis. Along the vertical axis – the flux density in arbitrary units (a.u.).}
    \label{fig:fig1}
\end{center}
\end{figure*}

\section{Observations and results}

After the reconstruction of the antenna, which took place in 2010-2012, the effective area of the LPA LPI increased to 45000 sq.m. in the direction of zenith. The geometric antenna area is 72000 sq.m. There was a fundamental possibility of implementing four independent radio telescopes based on one antenna field. New digital recorders have been made. For more information about scientific programs and details of radio telescope reconstruction, see \citeauthor{Shishov2016} (\citeyear{Shishov2016}) and \citeauthor{Tyulbashev2016} (\citeyear{Tyulbashev2016}).

A survey on one of the implementations of the LPA LPI radio telescope was carried out around the clock. 96 beams located in the plane of the meridian and fixed in the direction of the sky were used. The size of the radiation pattern of one beam is approximately $0.5^o \times 1^o$. The central frequency of observations is about 110.5 MHz with a full reception band of 2.5 MHz. The band is divided into separate, narrower frequency channels. Two recording modes are simultaneously implemented on the recorder: in one mode, the band is divided into 6 channels with width of 430 kHz and the readout time is 0.1s, and in the second, the band is divided into 32 channels of 78 kHz and the readout time is 12.5 ms.

The search for periodic radio emission from J0357+3205 was carried out in individual power spectra for all processed days using the BSA-Analytics program (https://github.com/vtyulb/BSA-analytics ; Qt/C++ under the GPL V3.0 license). At the first stage, the program was used to select power spectra having harmonics corresponding to the period J0357+3205, the signal-to-noise ratio of which was S/N > 3. There were 36 such spectra. At the second stage, days with raw data were taken and a pulsar search was carried out. The "single\textunderscore period" mode was used in the BSA-Analytics program. In this mode, the known period and coordinates of the pulsar are set, no sorting of periods and time shifts is done to identify the exact location of the pulsar in right ascension. The search is carried out at a recording length of 180s (± 1.5 minutes from the maximum of the LPA LPI radiation pattern). As a known parameter, a period of 0.4441 seconds was taken, and then the dispersion measures were sorted over the interval 0<DM<200 $pc/cm^3$. The main detection criterion for a pulsar with a known period was the appearance of a second peak on the average profile when the original signal with a double period was added together at the maximum of the S/N ratio on measures of dispersion close to the true one.

\begin{figure*}[h]
\begin{center}
	\includegraphics[width=1.0\textwidth]{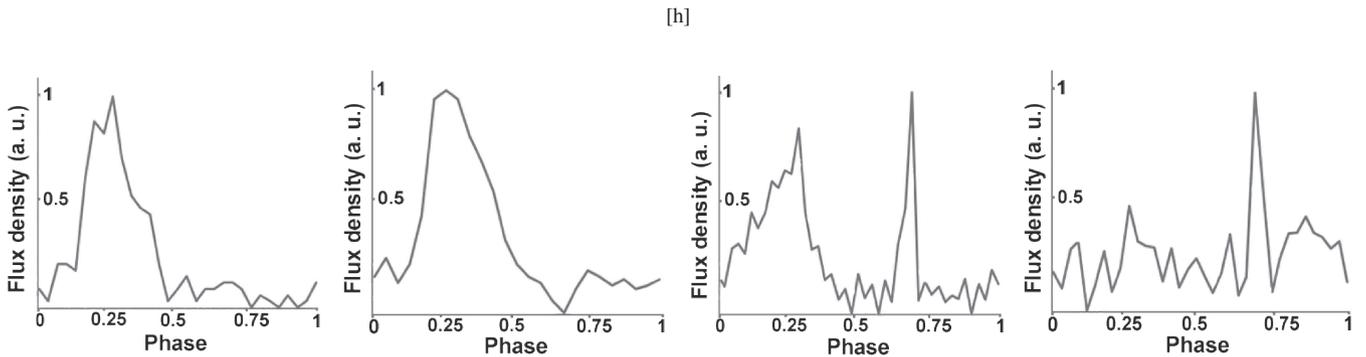}
    \caption{Contours of the average profiles of the gamma pulsar J0357+3205 obtained by addition with one period. From left to right are the contours extracted from the pictures of the average profiles in the gamma, X-ray and radio bands (\citeauthor{Abdo2009} (\citeyear{Abdo2009}), \citeauthor{Marelli2013} (\citeyear{Marelli2013}), \citeauthor{PeiWang2018} (\citeyear{PeiWang2018})). The rightmost figure is the average profile obtained from observations at the LPI LPA when averaging with a period P= 0.4441c. The horizontal axis in the figure shows the phase of the arrival of emission in the average profile in fractions of the period, the vertical axis shows the flux density in arbitrary units (a.u.).}
    \label{fig:fig2}
\end{center}
\end{figure*}

For two days (11/19/2015; 12/23/2016), peaks appeared in the region of DM=47 $pc/cm^3$ out of 36 processed on the "signal\textunderscore to\textunderscore noise dependence on the DM". The same DM was obtained from FAST observations (\citeauthor{PeiWang2018}, \citeyear{PeiWang2018}). Fig.~\ref{fig:fig1} shows the dependence of  S/N on DM and the profile accumulated with a double period on the DM = 47 $pc/cm^3$.

Let's compare our results with the results obtained by other authors. In the gamma range, one component is visible in the profile, which occupies a quarter of the period at half power (\citeauthor{Abdo2009}, \citeyear{Abdo2009}). That is, the half-width of the profile is about 100 ms. One component is also visible in the X-ray range (\citeauthor{Marelli2013}, \citeyear{Marelli2013}). Its half-width is also about 100 ms. In the radio range (\citeauthor{PeiWang2018}, \citeyear{PeiWang2018}), the profile is two-component. Based on the profile in the figure, one of the components is wide, and its half-width is approximately 90 ms. The second component is noticeably narrower, and is comparable in height to the wide component. Its half-width is approximately 25-30 ms. The distance between the components is approximately 165 ms. In our observations, only one narrow component is visible, and its half-width is approximately 20-25 ms, which is comparable to the width of the narrow component observed in FAST at a frequency of 1250 MHz. The height of the peaks in the average profile is 7 $\sigma_{noise}$. Unfortunately, in the conference presentation of FAST observations, it is not possible to compare the arrival phases of the components of the average profiles observed in the gamma, X-ray and radio bands. It can be assumed that the wide component observed in the gamma, X–ray and radio band (FAST) is the same, and the narrow component observed in FAST coincides with the narrow component observed on the LPA. Figure 2 shows the average profiles of observations in different ranges. The normalization of the profiles was made so that the maxima on the average profiles coincided in height. The lengths of the profiles were made the same.

Assuming that the pulsar has been detected, we will make a rough estimation of the peak pulsar flux density based on the theoretical estimate of the best sensitivity of the LPA LPI in the zenith direction (4.4 mJy; \citeauthor{Tyulbashev2016} (\citeyear{Tyulbashev2016})). Since the exact coordinates of the pulsar in right ascension and declination are known, it is possible to make corrections that take into account the features of the LPA LPI as a diffraction array. These corrections take into account the actual zenith distance of the source (correction =0.76), as well as the fact that the LPA LPI has a fixed direction of beams in the sky, does not coincide with the direction to the pulsar, and that the beam has a complex shape (correction= 0.47). We also take into account that S/N=10 in the resulting average profile (Fig.~\ref{fig:fig2}), and not S/N=6, as assumed in theoretical calculations (\citeauthor{Tyulbashev2016}, \citeyear{Tyulbashev2016}) (correction=10/6=1.67). The real pulse width is 5\%, not 10\%, as was thought for a typical pulse in the work [6] (correction= 0.70). Rough estimate of the flux density: $S_{111 MHz}$ =(4.4x1.67x0.70)/(0.76x0.47)=14 mJy.

A preliminary estimate of the observed pulsar flux density is 40 micro Jy at a frequency of 1250 MHz (\citeauthor{PeiWang2018}, \citeyear{PeiWang2018}). It is possible to make a formal estimation of the steepness of the spectrum $\alpha$, where $S \sim v ^{-\alpha}$. Spectral index is $\alpha$ =+2.4. The value of the spectral index may be significantly less, since periodic emission is found close to the sensitivity limit of the LPA LPI radio telescope and only once during all observation sessions.

% Don't change these lines
\bsp	% typesetting comment
\label{lastpage}
\end{document}